\documentclass[final,5p,times,twocolumn]{elsarticle} 
\DeclareMathAlphabet{\mathpzc}{OT1}{pzc}{m}{it}
\usepackage{lineno}
\usepackage{setspace}
\usepackage{graphicx}
\usepackage{amssymb}
\usepackage{amsmath}
\usepackage{amsthm}
\usepackage[colorlinks=true]{hyperref}
\usepackage[all]{hypcap}
\usepackage{mathrsfs}
\usepackage[percent]{overpic}
\usepackage{fixltx2e}
\usepackage{epsfig}
\usepackage{verbatim}
\usepackage{multirow}
\usepackage{float}
\usepackage{array}
\usepackage{alltt}
\usepackage{relsize}
\usepackage{upgreek}
\usepackage{arydshln}

\newcommand{\ngam}{N_0} 
\newcommand{\ntot}{N_{\alpha\beta}}
\newcommand{\var}{\mathrm{Var}}
\newcommand{\A}{\mathcal{A}^{(s)}}
\newcommand{\V}{\mathbf{V}}
\newcommand{\W}{\mathbf{W}}
\newcommand{\M}{\mathbf{M}}
\newcommand{\x}{\mu}

\newcommand{\pull}{}

\journal{Nuclear Instruments and Methods A}
\begin{document}
\begin{frontmatter}

\title{A synchronization method for the multi-channel silicon telescope}

\author[a]{P.~\v{Z}ugec\corref{cor1}}\ead{pzugec@phy.hr}
\author[b,c]{M.~Barbagallo}
\author[d]{J.~Andrzejewski}
\author[d]{J.~Perkowski}
\author[b]{N.~Colonna}
\author[a]{D.~Bosnar}
\author[d]{A.~Gawlik}
\author[c,e]{M.~Sabat\'{e}-Gilarte}
\author[c,f]{M.~Bacak}
\author[c]{F.~Mingrone}
\author[c]{E.~Chiaveri}

\address[a]{Department of Physics, Faculty of Science, University of Zagreb, Croatia}
\address[b]{Istituto Nazionale di Fisica Nucleare, Sezione di Bari, Italy}
\address[c]{European Organization for Nuclear Research (CERN), Geneva, Switzerland}
\address[d]{Uniwersytet \L\'{o}dzki, Lodz, Poland}
\address[e]{Universidad de Sevilla, Spain}
\address[f]{Technische Universit\"{a}t Wien, Vienna, Austria}

\author{\linebreak The n\_TOF Collaboration\fnref{fn1}} 
\cortext[cor1]{Corresponding author. Tel.: +385 1 4605552}
\fntext[fn1]{www.cern.ch/ntof}

\begin{abstract}
A simple method is presented for the simultaneous off-line synchronization of the digitally recorded data-streams from a multi-channel silicon telescope. The method is based both on the synchronization between the separate pairs of silicon strips and on the synchronization relative to an external timing device. Though only a reduced subset of these constraints is necessary in ideal circumstances, it is shown that this minimal set of conditions may not be sufficient for adequate synchronization in all cases. All available sources of information are therefore considered, in order to constrain the final synchronization as well as possible.
\end{abstract}

\begin{keyword}
Silicon telescope
\sep
Multi-channel synchronization
\sep
Neutron time of flight
\sep
n\_TOF facility
\end{keyword}
\end{frontmatter}

\section{Introduction}
\label{introduction}

The synchronization between multiple sampling channels is a common enough challenge in experimental nuclear physics, as well as other areas of research and technology. To this end, many different solutions were developed (see, for example, Refs.~\cite{sync1,sync2,sync3,sync4,sync5}). Older data acquisition systems, relying on the analogue electronic units such as the time-to-digital converters (TDC) and signal discriminators, have to be synchronized in advance, by a careful adjustment of the delay lines and the signal intake settings. The more recent types of digital electronics, such as the fast signal digitizers, profit from the possibility of implementing the complex on-the-fly or post-processing synchronization algorithms (to be applied during or after the signal acquisition). Naturally, an absolute time calibration requires a timing reference, typically an external timing device (see Ref.~\cite{sync3} for a concise and succinct description). Implementation of such on-the-fly algorithms is, of course, more challenging than of the post-processing ones, as the additional hardware and signal interlacing requirements need to be met. We provide here a simple, purely post-processing method that can be applied after the pulse-processing stage of extracting the physical data from the registered signals. The obvious practical advantage of such \textit{a posteriori} method is that it can be utilized at the very late stage in the data analysis, without having to reprocess the signals in case the time offsets between multiple channels were belatedly identified. In that, the method simultaneously takes into account both the absolute timing constraints -- in respect to the external timing device -- and the relative timing constraints between all admissible pairs of channels. This feat is based on observing the statistical properties of already identified pulses, which could hardly be achieved by other means.

Section~\ref{experimental} describes the details of the experimental setup and the context of the synchronization issues. Section~\ref{method} presents the proposed synchronization method: all the necessary considerations to be taken into account, as well as the necessary implementation details. Section~\ref{conclusions} summarizes the main conclusions of this work.

\section{Experimental setup}
\label{experimental}

The neutron time of flight facility n\_TOF at CERN is the highly luminous white neutron source spanning 12 orders of magnitude in neutron energy -- from 10~meV to 10~GeV. Its operation is based on the 20 GeV proton beam from the CERN Proton Synchrotron irradiating a massive lead spallation target, serving both as the neutron source and the primary moderator of the initially fast neutrons. The second stage of moderation takes place in the borated or demineralized layer of water from the cooling system surrounding the spallation target. The general features of the n\_TOF facilty are well documented and may be found in Ref.~\cite{ntof}.

Today the n\_TOF facility accommodates two experimental areas: Experimental Area 1 (EAR1) located at the horizontal distance of 185~m from the spallation target \cite{ntof} and the Experimental Area 2 (EAR2) situated 20~m above the same target \cite{ear2_1,ear2_2,ear2_3}. Each experimental area is specially suited to the particular set of challenges in measuring the different types of neutron induced reactions: from the neutron capture and the neutron induced fission to the reactions with the light charged particles in the exit channel \cite{ntof_rev}. EAR1 offers an excellent neutron energy resolution and allows for the high neutron energy measurements due to the increased neutron flight path. Thanks to the extremely high instantaneous neutron flux EAR2 provides the unprecedented capabilities for measuring very low neutron reaction cross sections, including the measurements with very small and/or highly radioactive samples.

\begin{figure}[t!]
\centering
\includegraphics[width=0.9\linewidth]{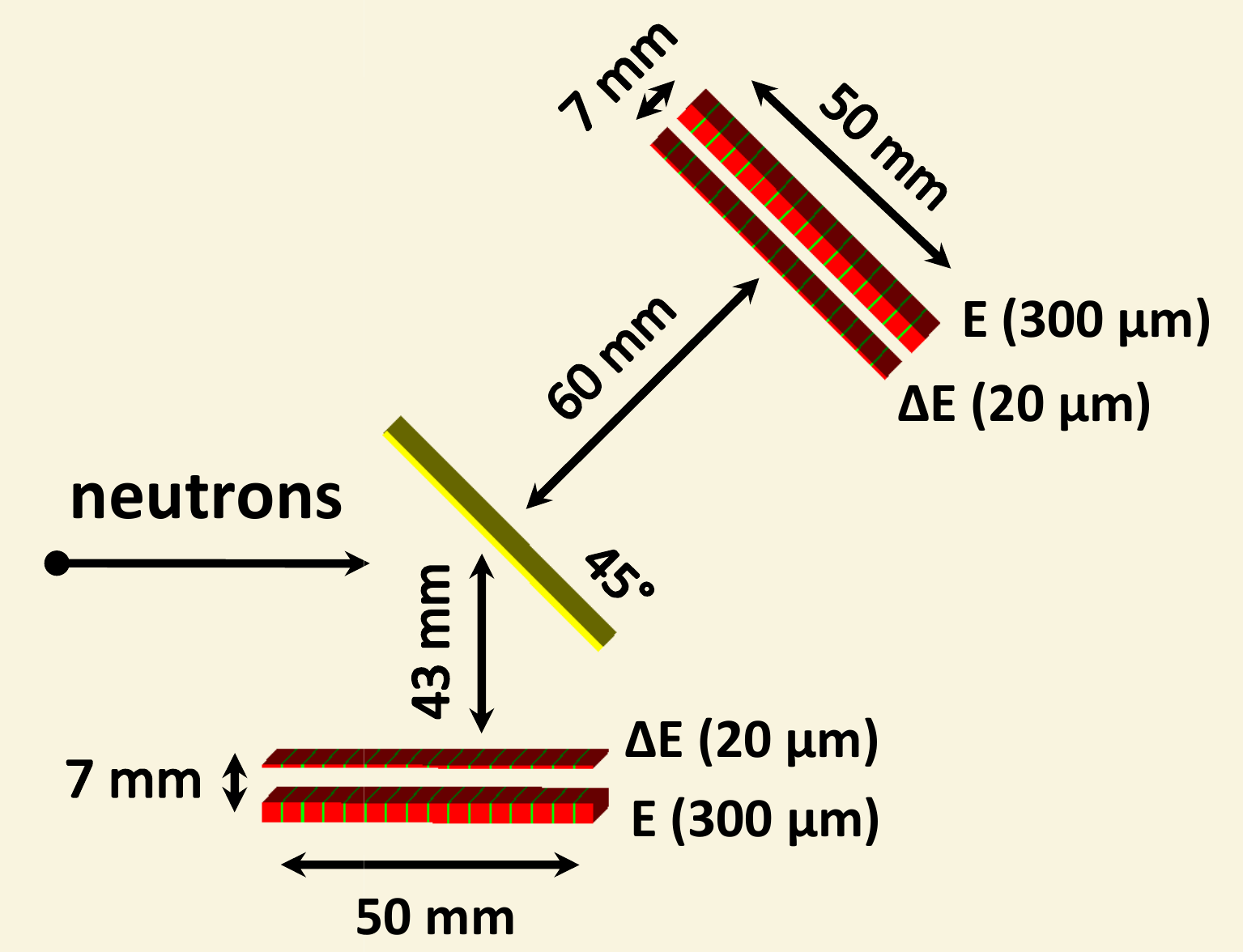}
\includegraphics[width=0.9\linewidth]{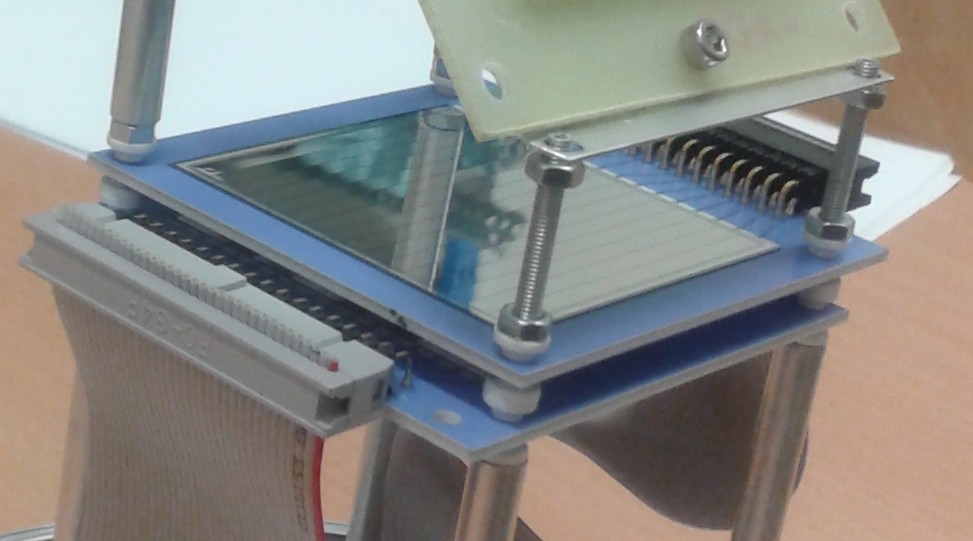}
\caption{Top: detector configuration from the joint measurement of the  $^{12}$C($n,p$) and $^{12}$C($n,d$) reactions, comprising two identical silicon telescopes (all widths are exaggerated). At the center is the bearing structure for the LiF calibration sample (we use these calibration data in this work). Bottom: SITE close-up. The striped structure of $\Delta E$-layer (shown here) is clearly visible.}
\label{fig1}
\end{figure}

Many different types of detectors are employed at n\_TOF, each specially suited to the measurements of the particular type of reaction. A general review of these detectors and the associated signal analysis procedures may be found in Ref.~\cite{psa}. One of these detectors -- relevant to this work -- is the multi-channel silicon telescope (SITE), recently introduced at n\_TOF for measurements of the neutron induced reactions with the light charged particles in the exit channel, such as the ($n,p$), ($n,d$), ($n,t$), ($n,\alpha$) reactions \cite{site_np}. This detector was first used in the highly challenging measurement of the $^7$Be($n,p$) reaction \cite{be_np}, which was also accompanied by the measurement of the $^7$Be($n,\alpha$) reaction \cite{be_na}, relying on the similar type of the silicon sandwich detector \cite{sandw_na}. Both of these measurements, of central importance for the famous and as yet unresolved Cosmological Lithium Problem, became feasible only with the successful construction of EAR2.

Two such multi-channel telescopes were recently used in the joint energy-differential measurement of the $^{12}$C($n,p$) and $^{12}$C($n,d$) reactions \cite{carbon_prop}, performed at EAR1 of the n\_TOF facility. This measurement was motivated by the unexpected results from an earlier integral measurement of the $^{12}$C($n,p$) reaction \cite{carbon_prc,carbon_epja}, yielding an integral cross section higher than indicated by any past dataset. The analysis of the data from the latest $^{12}$C($n,p$) and $^{12}$C($n,d$) measurements is under way, and the special analysis procedure has already been developed in order to properly take into account the challenging nature of these reactions \cite{angular}. The geometric configuration of the used detector setup is shown in Fig.~\ref{fig1}. We refer to the top telescope as the \textit{front} SITE and the bottom one as the \textit{rear} SITE. In short, each SITE consists of two segmented layers of silicon strips, each layer comprising 16 strips, all of them oriented in the same direction (rather than perpendicular between the layers). Both layers are 5~cm~$\times$~5~cm, with active strips of 5~cm~$\times$~3~mm separated by a thin layer of inactive silicon. The two layers -- the first, $\Delta E$-layer and the second, $E$-layer -- are distanced by 7~mm. Their respective thicknesses are 20~$\mu$m and 300~$\mu$m. Further details about the telescope construction and readout may be found in Ref.~\cite{site_np}.

\begin{figure}[t!]
\centering
\includegraphics[width=0.9\linewidth]{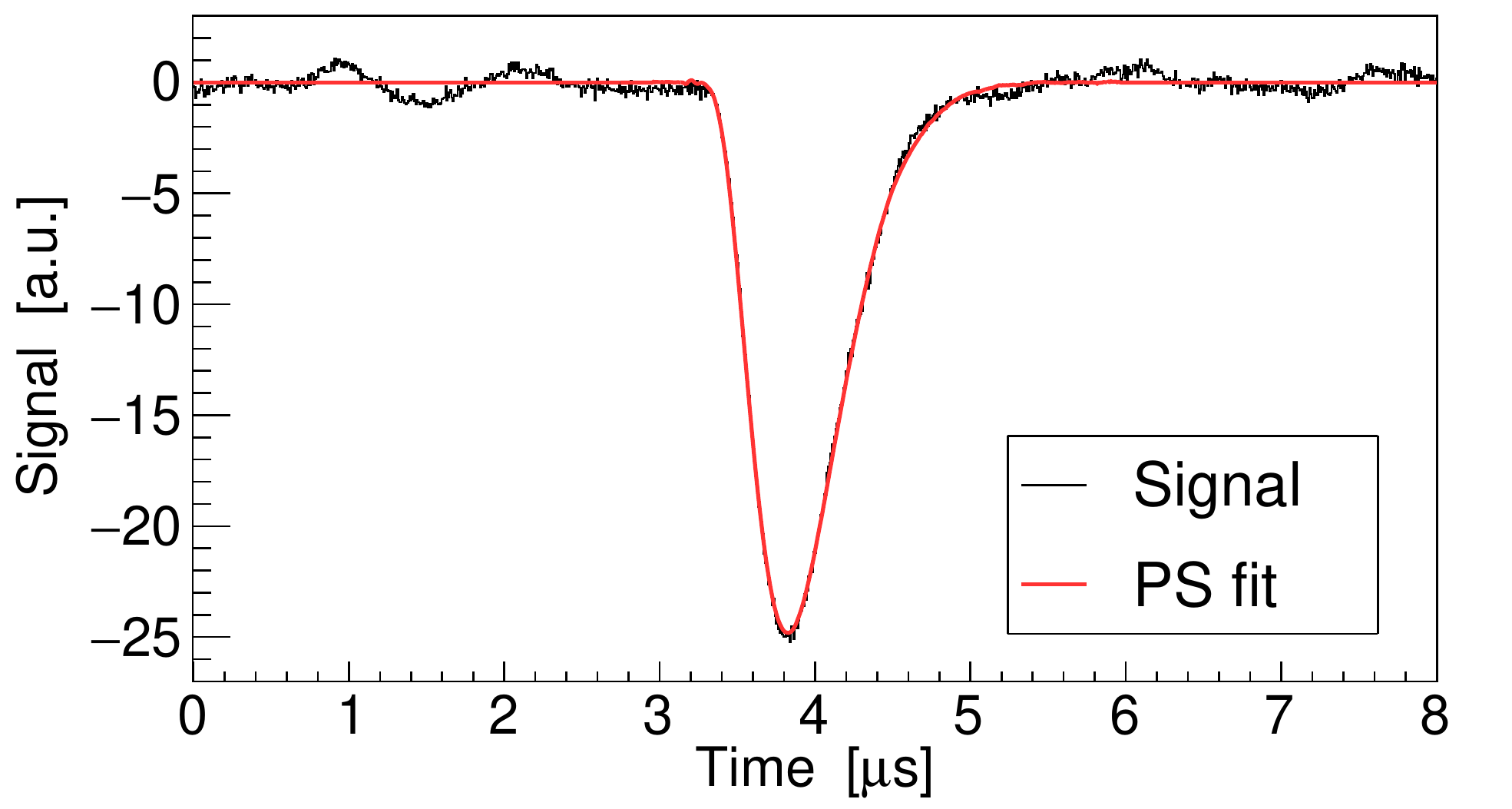}
\caption{Pulse from one of the SITE strips and the optimally adjusted numerical pulse shape. The amplitude and the timing properties are determined from the fit.}
\label{fig_psa}
\end{figure}

The signals from two SITE were recorded and digitized at 125~MS/s sampling rate with a 14-bit resolution. They were analyzed by the pulse shape fitting procedure described in Ref.~\cite{psa} -- specifically, by adjusting a numerical pulse shape to the baseline-corrected pulses and extracting the pulse amplitude and timing properties from such fit. An example of the SITE pulse, together with the adjusted pulse shape is shown in Fig.~\ref{fig_psa}.

Another detector of importance to this work is the Wall Current Monitor (WCM) \cite{pkup} -- an induction device specifically designed for registering the proton pulses delivered by the Proton Synchrotron. WCM offers a reliable response to a proton pulse, registering with high fidelity the intensity of the beam, as well as the arrival time of the pulse.

The main purpose of this work is to provide a simple method for a simultaneous synchronization of the digitally recorded data-streams from all involved detector channels, i.e. from all silicon strips. The proper temporal synchronization of all channels is crucial for the proper identification of the coincidental pulses between the two ($\Delta E$ and $E$) layers, signaling the detection of a charged particle. The basic idea has already been used to determine the synchronization between the entire $\Delta E$-layer and entire $E$-layer in measurement with the $^7$Be sample \cite{site_np}, and to determine the appropriate coincidence window width for the $^7$Be($n,p$) data. However, no notable time offsets within a given layer were observed, so it was sufficient to consider only a single spectrum of time differences between the pulses from any silicon strip in $\Delta E$-layer and any strip in $E$-layer (the \textit{overall} distribution of differences), immediately yielding the data-recording time offset between the two layers as the mean value of this distribution. During the joint $^{12}$C($n,p$) and $^{12}$C($n,d$) measurement the time offsets between several strips from within the same layer were observed. The reason may be as simple as using transmission lines of mismatched length in transmitting the signals from the particular strips to the digital data recording system. There may also be other, unaccounted sources of time offset within the entire data acquisition chain. In regard to the $^7$Be measurement, it must be taken into account that a different acquisition chain was used -- the one dedicated to EAR2, as opposed to the one from EAR1 in case of the $^{12}$C measurement -- thus justifying the difference in the synchronization issues between the two experiments. Therefore, we expand here the basic idea from Ref.~\cite{site_np}, providing the general procedure for the simultaneous synchronization of all data channels, taking into account any and all timing information available in order to constrain the time offsets between the particular strips and some external clock as well as possible.

For purposes of energy calibration of the silicon strips, a measurement of the $^6$Li($n,t$) reaction was performed during the $^{12}$C campaign, using the $^6$Li-enriched LiF calibration sample. The synchronization issues, i.e. the time offsets between multiple channels are, of course, independent of the used sample and/or the measured reaction. In this work we show the data from the $^6$Li($n,t$) measurement for a simple reason: they yield slightly more presentable plots. In addition, if the synchronization procedure needs to be repeated in a course of some other campaign, the main reaction of interest will change. On the other hand, the $^6$Li($n,t$) measurement is a standard calibration procedure employed at n\_TOF, so that by presenting these particular data we provide the possibility for the fully consistent check against the measurement that is repeated between different campaigns.

Incidentally, the time offsets -- which, in our case, are mostly contained below 100~ns -- would not, if left uncorrected, notably affect the time of flight spectra of the $^{12}$C($n,p$) and $^{12}$C($n,d$) reactions, as their relevant energy range around 20~MeV (aimed by the measurement from n\_TOF) corresponds to EAR1 time of flight to of approximately 3~$\mu$s. However, the identification of coincidences between the strips does critically depend on the adequate synchronization, as the offsets may become comparable to or larger than the appropriate coincidence windows. In addition, if the reactions of interest were taking place at higher neutron energies, or even if the intermediate-energy measurements were performed at shorter flight paths -- such as EAR2 of the n\_TOF facility -- the measured time of flight spectra might become severely affected even by the time offsets in excess of tens of nanoseconds.

\section{Synchronization method}
\label{method}

Let $n$ enumerate the relevant events, be it the instance of some reaction of interest or some separate event. Let $s$ enumerate the particular silicon strips (of which there are 32 in a single SITE from n\_TOF). Then the time instant $t_{ns}$ of the $n$-th event, \textit{as registered} by the $s$-th strip may be expressed as:
\begin{linenomath}\begin{equation}
t_{ns}\simeq T_n+\tau_s,
\label{master}
\end{equation}\end{linenomath}
with $T_n$ as some reference time of that event, and $\tau_s$ as this particular strip's offset relative to the timing device yielding $T_n$. For simplicity of notation we use throughout this paper the symbol $\simeq$ (to be read as "is expected to be") to indicate the statistically expected values, in a sense that \mbox{$\langle t_{ns}\rangle= \langle T_n+\tau_s\rangle$}. It is our goal to determine the offsets $\tau_s$, taking into account the maximum amount of information available from the experimental data, thus constraining the set of $\tau_s$ as well as possible.

There are two types of pulses regularly recorded by any detection system adopted at n\_TOF. One is the so-called $\gamma$-flash pulse, caused by an intense burst of $\gamma$-rays and ultrarelativistic particles from the spallation process producing the neutron beam. The other type consists of the those pulses following the $\gamma$-flash, related to the detection of the neutron-induced reactions of interest, and to the background processes caused by the competing neutron reactions and the environmental radioactivity. We will use the first type -- the $\gamma$-flash pulses -- for the absolute timing calibration, together with the second type -- the coincidental pulses from the detection of the neutron-induced reactions -- for the relative calibration of the strip offsets $\tau_s$.

\subsection{Absolute calibration}
\label{absolute}

For the reference time instants $T_n$ of the particular $\gamma$-flashes one could consider the $\gamma$-flash pulses from one selected silicon strip. However, due to the insensitivity-by-design (in order for a detector not to be blinded by the $\gamma$-flash) individual strips have a low efficiency for detecting the $\gamma$-flash. This is also the reason why, following each $\gamma$-flash, no strip can be consistently calibrated relative to its own $\gamma$-flash pulse, thus necessitating the external timing information. As an external timing device we use the WCM, whose reliable response to each and every proton pulse is immediately followed by the release of the $\gamma$-flash. Thus, we take the instant $T_n$ of the proton beam delivery, as registered by WCM, as the reference point for the absolute time calibration of silicon strips.

Let us define the time offset:
\begin{linenomath}\begin{equation}
\Delta_{ns}\equiv t_{ns}-T_n
\label{offset}
\end{equation}\end{linenomath}
for the $n$-th $\gamma$-flash pulse registered at $t_{ns}$ by the $s$-th strip. We will take into account the fidelity of the $\gamma$-flash detection by the particular strip (or its proper recognition during the pulse shape analysis of electronic signals) by weighting its contribution by the $\gamma$-flash pulse amplitude $A_{ns}$, as registered by that strip. The choice of the weighting factors -- these particular ones having been selected for the conceptual simplicity -- is essentially arbitrary, as the final results are rather insensitive to a wide class of alternative selections (e.g. $A_{ns}^{1/2}$), provided that the selected function of $A_{ns}$ (here linear) is neither pathological nor inordinately selective of any particular range of amplitudes. As opposed to the later unweighted procedure -- related to Eq.~(\ref{simple_mean}) and motivated therein by the amplitude variations reflecting an intrinsically meaningful spectrum -- weighting the strips' response to a $\gamma$-flash is justified by the registered $\gamma$-flash amplitudes indeed being a measure of the reliability of this response. The reason in twofold: (1)~the intensity of the proton beam from the Proton Synchrotron -- as the primary cause of the $\gamma$-flash -- is well defined and usually limited to one of the two particular values (the lower and higher intensity, known internally as the parasitic and dedicated mode), with very little variations, while there \textit{are} broad amplitude variations in each strip's response; (2)~there are broad variations in different strips' response to the \textit{same} $\gamma$-flash.

For those strips that did not register the $\gamma$-flash, or the registered pulse was rejected for any reason during the data analysis, one simply takes \mbox{$A_{ns}=0$} as the weighting factor. Defining, for convenience of notation:
\begin{linenomath}\begin{equation}
\textstyle \A_1\equiv\sum_{n=1}^{\ngam} A_{ns} \quad\text{and}\quad \A_2\equiv\sum_{n=1}^{\ngam} A_{ns}^2,
\end{equation}\end{linenomath}
with $N_0$ as the total number of proton pulses delivered, we may express the weighted averaged offset for the $s$-th strip as:
\begin{linenomath}\begin{equation}
\bar{\Delta}_s\equiv \frac{\sum_{n=1}^{\ngam} A_{ns}\Delta_{ns}}{\A_1},
\end{equation}\end{linenomath}
together with the unbiased estimator of its weighted variance:
\begin{linenomath}\begin{equation}
\var_s\,\Delta=\frac{\left(\sum_{n=1}^{\ngam} A_{ns}\Delta_{ns}^2\right)-\A_1\bar{\Delta}_s^2}{\A_1-\A_2/\A_1}.
\label{var_big}
\end{equation}\end{linenomath}
In that, $\var_s\,\Delta$ is the sample variance (related to the width of the distribution of $\Delta_{ns}$), as opposed to the variance of the mean $\bar{\Delta}_s$, which equals: \mbox{$\var_s\,\bar{\Delta}=\var_s\,\Delta\times\A_2/\big(\A_1\big)^2$}.

\begin{figure}[t!]
\centering
\includegraphics[width=1\linewidth]{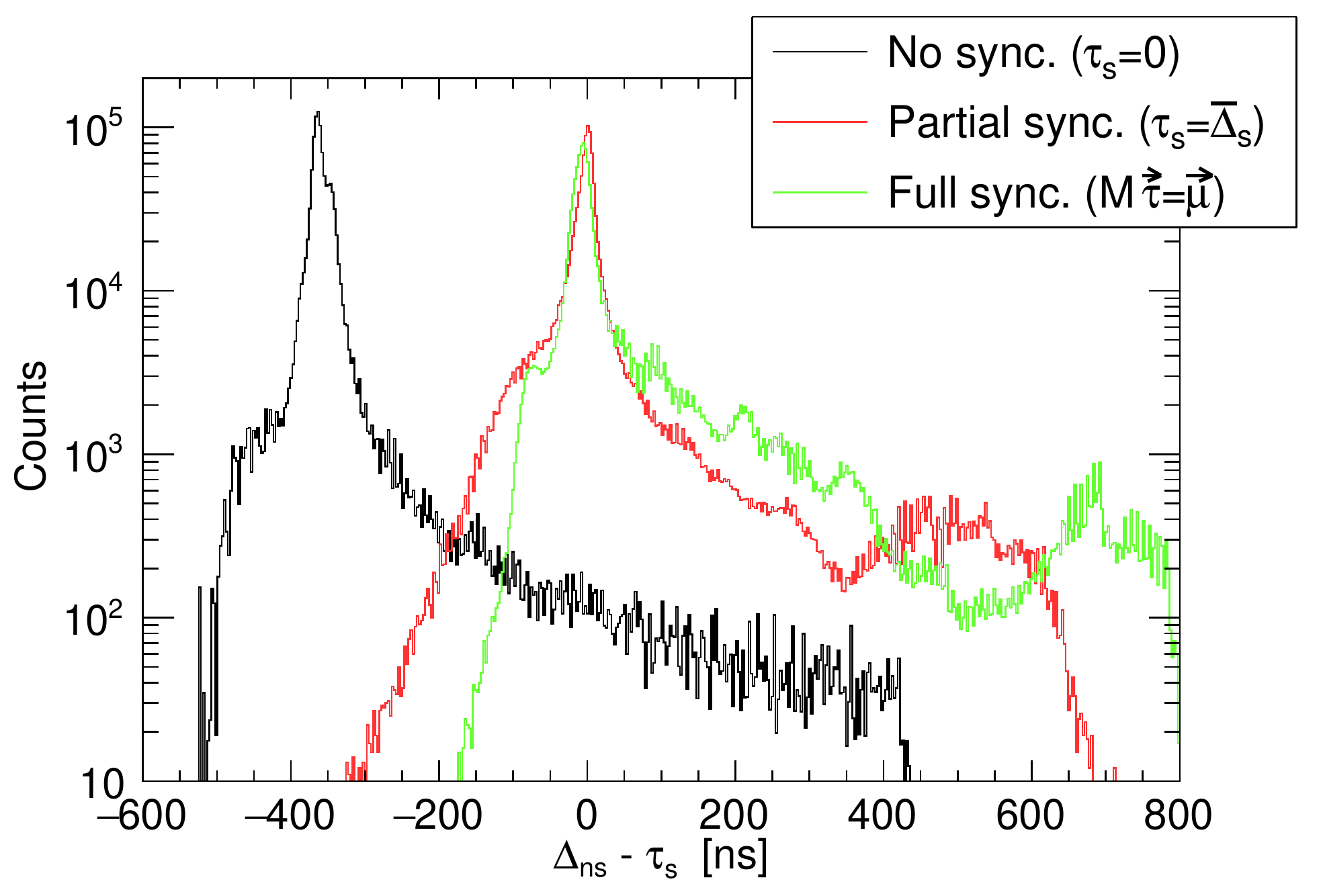}
\pull
\caption{Overall distribution of differences between the $\gamma$-flash instants as registered by a particular silicon strip and as registered by an external timing device (WCM). The synchronization procedure aims at identifying the set of offsets $\tau_s$ such that the condition from Eq.~(\ref{abs_con}) is satisfied, i.e. that the mean of the corrected spectra is as close to zero as possible. The full synchronization procedure takes into account further requirements from Eqs.~(\ref{rel_con}) and (\ref{add_con}).}
\label{fig2}
\end{figure}

The black plot from Fig.~\ref{fig2} shows a total distribution of offsets $\Delta_{ns}$ from all 64 silicon strips, related to the $^6$Li($n,t$) calibration measurement. One can clearly observe that the average offset between the WCM and SITE signals is approximately 350~ns. The root-mean-square (RMS) of the distributions for particular strips varies between 30~ns and 280~ns. The rest of the plots from Fig.~\ref{fig2} will be discussed later.

From the definition of offsets $\Delta_{ns}$ by Eq.~(\ref{offset}) and the central relation from Eq.~(\ref{master}) it is evident that the mean offsets $\bar{\Delta}_s$ provide a set of estimators for the sought offsets $\tau_s$:
\begin{linenomath}\begin{equation}
\tau_s \simeq \bar{\Delta}_s.
\label{abs_con}
\end{equation}\end{linenomath}
This is a first set of equations for $\tau_s$, one that apparently immediately provides the entire set of $\tau_s$. However, we shall soon observe that this set of conditions is of insufficient quality. The reason is precisely the occasional inaccuracy in the $\gamma$-flash pulse recognition within some of the silicon strips. In the presence of competing pulses in the immediate vicinity of the supposed $\gamma$-flash pulse, an erroneous pulse may be assigned a status of a $\gamma$-flash pulse. These sporadic misidentifications propagate into the calculation of the average $\bar{\Delta}_s$, thus making these estimators prone to a certain degree of error. We will therefore use additional sources of information. In conjunction with these, the conditions from Eq.~(\ref{abs_con}) will be shown to behave only as the good initial estimates for $\tau_s$.

\subsection{Relative calibration}
\label{relative}

The additional constraints upon the set of offsets $\tau_s$ may be obtained by observing the time differences between the pairs of pulses from any pair of strips. From entirely uncorrelated pulses one would expect a flat or featureless contribution to the spectrum of time differences. On the other hand, coincidental pulses gather around a well defined value, corresponding to a relative offset between the strips, thus forming a recognizable spectral peak.

Let $\alpha$ and $\beta$ denote the two strips either from the same or separate ($\Delta E$ or $E$) silicon layer. We now consider the time differences between all the pulses registered during the particular measurement. In this case the index $n$ from Eq.~(\ref{master}) denotes all detected counts -- as opposed to the sole $\gamma$-flash pulses from Section~\ref{absolute} -- while $T_n$ corresponds to an instant of the $n$-th detected event as it would have been registered by an external timing device (WCM) if this device were used for the detection of these events. Simply put, it is the detection time according to an external clock. Defining the time difference between the coincidental counts from strip $\alpha$ and strip $\beta$:
\begin{linenomath}\begin{equation}
\delta_{n\alpha\beta}\equiv t_{n\alpha}-t_{n\beta},
\label{rel_diff}
\end{equation}\end{linenomath}
one immediately observes that it will be invariant of $T_n$ and will provide an estimator for the relative offset \mbox{$\tau_\alpha-\tau_\beta$} between the two strips.

\begin{figure}[t!]
\centering
\includegraphics[width=1\linewidth]{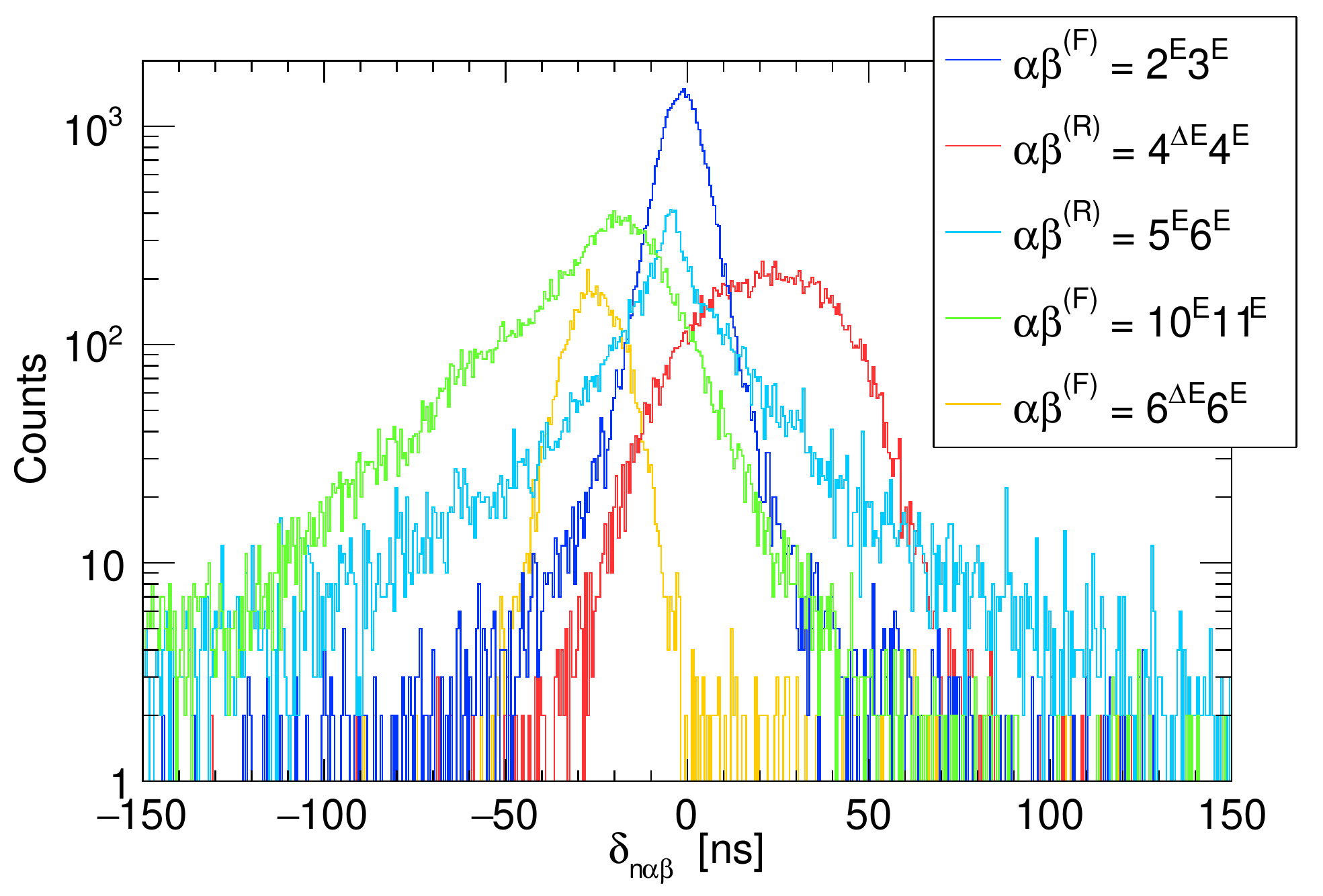}
\pull
\caption{Distribution of the registered-time differences between the counts detected in coincidence by several arbitrarily selected ($\alpha,\beta$)-pairs of strips either from the same ($\Delta E$ or $E$) or separate ($\Delta E$ and $E$) silicon layers. In legend the designation F/R indicates either the front (F) or rear (R) telescope (see the top scheme from Fig.~\ref{fig1}). Numbers are strip designations from a given layer (1--16), with the layer itself being identified by the superscript $\Delta E$ or $E$.}
\label{fig3}
\end{figure}

It should be noted that by plotting the distribution of time differences between \textit{any and all} pairs of pulses from a particular pair of strips, one in general observes the differences \mbox{$t_{n\alpha}-t_{m\beta}$} between the separate ($n$-th and $m$-th) events. However, by recognizing and selecting only the counts from the coincidental spectral peak, one ensures that both time instants from Eq.~(\ref{rel_diff}) belong to the same ($n$-th) event. Of course, when there are reasonable indications for the expected relative offset \mbox{$\tau_\alpha-\tau_\beta$}, one does not need to consider all the possible pairs of pulses from the separate strips. Otherwise, the procedure is apt to become extremely computationally inefficient, especially when the recorded data-stream (signal waveform) is much longer than the expected offset between the strips, which is certainly the case at n\_TOF ($\sim$100~ms waveform vs. \mbox{$\max_{\alpha\beta}|\tau_\alpha-\tau_\beta|\approx$100~ns}). Instead, one just considers the pairs of pulses within the appropriate time window. Alongside the response to the $\gamma$-flash, the physical cause for the coincidences between the strips from separate ($\Delta E$ and $E$) silicon layers is self-evident, as it constitutes the working principle of the silicon telescope -- it is the detection of a charged particle passing through both layers (tritons from the $^6$Li($n,t$) reaction in case of the energy calibration data used in this work). Moreover, even for the certain immediate pairs of strips from the \textit{same} silicon layer there are available coincidental counts, caused either by the $\gamma$-flash or by the signal separation between the neighboring strips due to the charge particle passing close to their shared boundary. This is certainly a source of information to be further exploited. Figure~\ref{fig3} shows the coincidental peaks in the distribution of time differences $\delta_{n\alpha\beta}$ for several arbitrarily selected $(\alpha,\beta)$-pairs of strips either from the same or separate silicon layers. Among the coincidences within the same layer, those from $E$-layer are more frequent than those from $\Delta E$-layer due to the thinner strips' increased insensitivity to the $\gamma$-flash.

Similarly to the calibration relative to the external timing device (WCM), we will consider the mean value $\bar{\delta}_{\alpha\beta}$ as the relevant estimator for the inter-strip offset. However, this time we adopt a simple unweighted mean:
\begin{linenomath}\begin{equation}
\bar{\delta}_{\alpha\beta}=\frac{\sum_{n=1}^{\ntot}\delta_{n\alpha\beta}}{\ntot},
\label{simple_mean}
\end{equation}\end{linenomath}
with $N_{\alpha\beta}$ as the total number of coincidental counts detected by the ($\alpha,\beta$)-pair of strips. This selection is motivated by the coincidental counts caused by the detection of charged particles produced in the sample. They are characterized by an extensive deposited-energy spectrum, affected by the intrinsic spectrum of the measured nuclear reaction(s) and by the interaction of charged particles with the silicon detector. The amplitude of these counts is thus a consistent and physically meaningful quantity, rather than the measure of the reliability of the detector response. As such, these amplitudes may no longer be considered as the weighting factors affecting the significance of particular inputs $\delta_{n\alpha\beta}$ to Eq.~(\ref{simple_mean}). In that, the unbiased estimator of the sample variance:
\begin{linenomath}\begin{equation}
\var_{\alpha\beta}\,\delta=\frac{\left(\sum_{n=1}^{\ntot}\delta_{n\alpha\beta}^2\right)-\ntot\bar{\delta}_{\alpha\beta}^2}{\ntot(\ntot-1)}
\label{var_small}
\end{equation}\end{linenomath}
is also directly related to the variance of the mean value from Eq.~(\ref{simple_mean}) as: \mbox{$\var_{\alpha\beta}\,\bar{\delta}=(\var_{\alpha\beta}\,\delta)/\ntot$}.

From Eqs.~(\ref{master}) and (\ref{rel_diff}) it is now evident that the averages $\bar{\delta}_{\alpha\beta}$ serve as the inter-strip offset estimators:
\begin{linenomath}\begin{equation}
\tau_\alpha-\tau_\beta \simeq \bar{\delta}_{\alpha\beta},
\label{rel_con}
\end{equation}\end{linenomath}
thus providing a second set of constraints upon the sought offsets $\tau_s$, in addition to the one from Eq.~(\ref{abs_con}). One is, of course, well advised to use only the constraints from those pairs of strips that feature a statistically significant number $N_{\alpha\beta}$ of coincidental counts and/or acceptably low variance $\var_{\alpha\beta}\,\bar{\delta}$ of the obtained mean values.

\subsection{Additional constraints}
\label{extra}

If necessary, one can also attempt to construct additional \textit{linearly independent} constraints, alongside those from Eqs.~(\ref{abs_con}) and (\ref{rel_con}). We illustrate here one such example that may be of further help in obtaining as accurate values of $\tau_s$ as possible.

\begin{figure}[t!]
\centering
\includegraphics[width=1\linewidth]{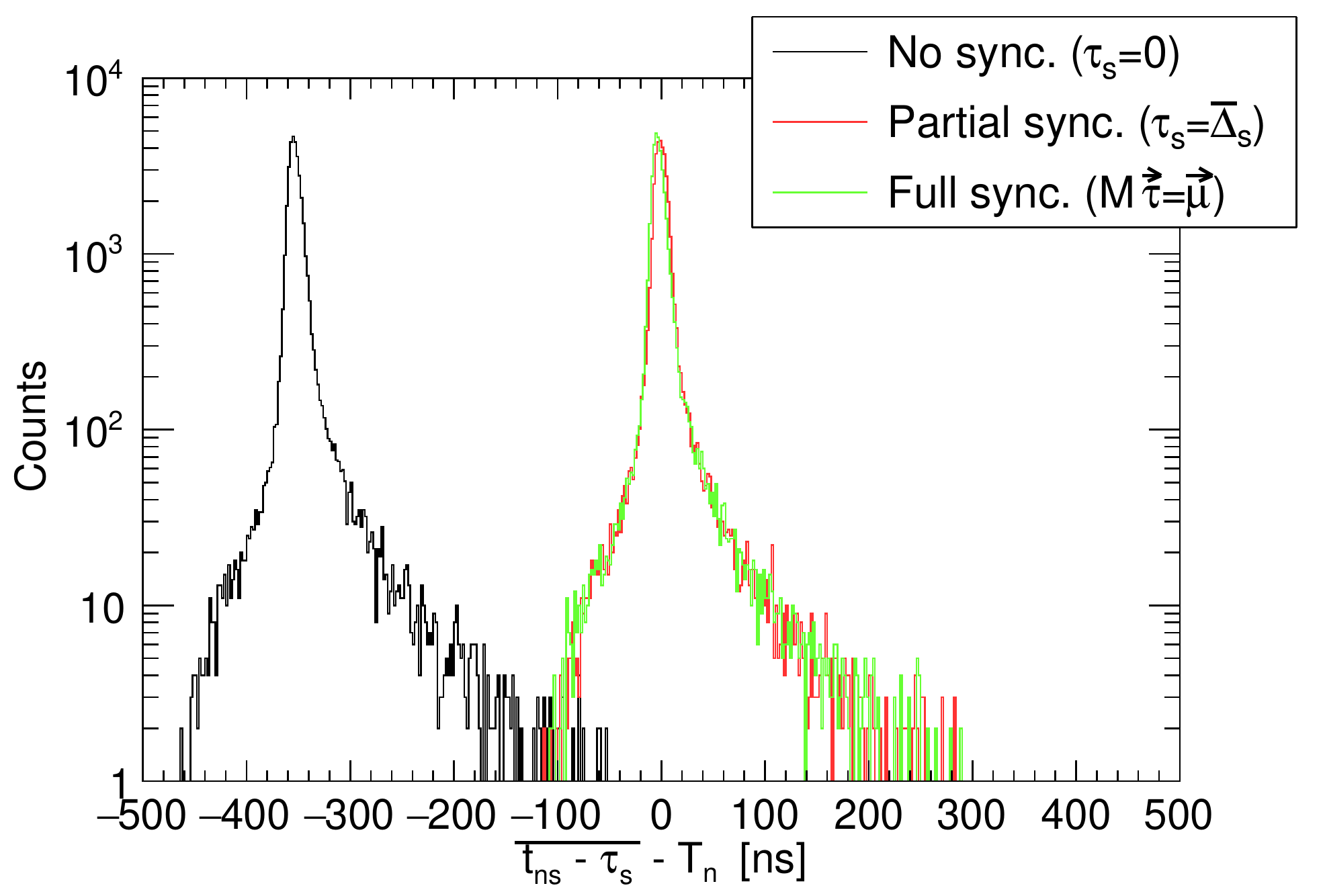}
\pull
\caption{Overall distribution of averaged $\gamma$-flash instants from Eq.~(\ref{mean_t}), as registered by separate silicon strips. Already the partially synchronized spectrum -- accounting only for the set of conditions from Eq.~(\ref{abs_con}) -- satisfies to a high degree the expectation from Eq.~(\ref{lin_ind}). Full synchronization considers this as an explicit requirement, alongside Eqs.~(\ref{abs_con}) and (\ref{rel_con}).}
\label{fig4}
\end{figure}

For the total of $S$ available silicon strips we define a weighted average $\bar{T}_n$ of already synchronized $\gamma$-flash instants, as registered by separate strips:
\begin{linenomath}\begin{equation}
\bar{T}_n\equiv \frac{\sum_{s=1}^S A_{ns}(t_{ns}-\tau_s)}{\sum_{s=1}^S A_{ns}}
\label{mean_t}
\end{equation}\end{linenomath}
and demand that their average deviation from the $\gamma$-flash instants registered by an external timing device (WCM) vanishes:
\begin{linenomath}\begin{equation}
\textstyle \sum_{n=1}^{\ngam}(\bar{T}_n-T_n)\simeq0,
\label{lin_ind}
\end{equation}\end{linenomath}
with $N_0$ as the total number of proton pulses, just as in Section~\ref{absolute}. The black plot from Fig.~\ref{fig4} shows a distribution of $\gamma$-flash instants without correction for time offsets, averaged over all available strips, i.e. the distribution of terms \mbox{$\sum_{s=1}^S A_{ns}t_{ns}\,\big/\sum_{s=1}^S A_{ns}$}, serving as the basis of Eq.~(\ref{mean_t}). The rest of the plots will be discussed later.

One needs to be mindful of the reliability of conditions like Eq.~(\ref{lin_ind}), since there are sporadic misidentifications in the $\gamma$-flash instants $t_{ns}$ extracted from particular strips (see Section~\ref{absolute}). However, a massively decreased width of the distributions from Fig.~\ref{fig4} (RMS of 22~ns), relative to the ones from Fig.~\ref{fig2} (RMS for particular strips from 30~ns to 280~ns) confirms that the misidentified pulses are of low amplitude. Thus, their contribution to the average $\bar{T}_n$ is heavily suppressed by the weighting procedure.

Equation~(\ref{lin_ind}) may now be rewritten as:
\begin{linenomath}\begin{equation}
\textstyle \sum_{s=1}^S w_s\tau_s \simeq \bar{\tau},
\label{add_con}
\end{equation}\end{linenomath}
with the following terms, easily obtained upon its careful rearrangement:
\begin{linenomath}\begin{align}
&w_s =\frac{1}{\ngam} \sum_{n=1}^{\ngam}\frac{A_{ns}}{\sum_{\sigma=1}^SA_{n\sigma}},\\
&\bar{\tau} =\frac{1}{\ngam}\sum_{n=1}^{\ngam}\frac{\sum_{\sigma=1}^SA_{n\sigma}\Delta_{n\sigma}}{\sum_{\sigma=1}^SA_{n\sigma}}.
\end{align}\end{linenomath}
Evidently, Eq.~(\ref{add_con}) is an additional constraint upon the set sought $\tau_s$, linearly independent of Eqs.~(\ref{abs_con}) and (\ref{rel_con}). While the independence from the set of of equations~(\ref{rel_con}) is self-evident, as they refer to a different dataset, the independence from the set of equations~(\ref{abs_con}) is demonstrated in \mbox{\ref{linear}}.

\subsection{Simultaneous synchronization}

By Eqs.~(\ref{abs_con}), (\ref{rel_con}) and (\ref{add_con}) we constructed an \textit{overdetermined} set of constraints for the set of sought offsets $\tau_s$. This entire system of linear equations may be put into a matrix form:
\begin{linenomath}\begin{equation}
\M\vec{\tau}=\vec{\x}
\label{vectorized}
\end{equation}\end{linenomath}
and it is easily solved in a least-squares sense, by weighted fitting. Let us illustrate the structure of the design matrix $\M$ on an artificial example of a single SITE comprising three strips in $\Delta E$-layer ($s=1,2,3$) and three strips in $E$-layer ($s=4,5,6$), while assuming that only the coincidences between the closest strips within the same layer are available, together with the closest and next-to-closest strips between separate layers:
\begin{linenomath}\begin{equation*}
\left[\begin{array}{rrr;{1pt/1.5pt}rrr}
1&0&0&0&0&0\\ 
0&1&0&0&0&0\\ 
0&0&1&0&0&0\\ 
0&0&0&1&0&0\\ 
0&0&0&0&1&0\\ 
0&0&0&0&0&1\\ 
\hdashline[1pt/1.5pt]\hdashline[1pt/1.5pt]&&&&&\\[-1em]
1&-1&0&0&0&0\\ 
0&1&-1&0&0&0\\ 
0&0&0&1&-1&0\\ 
0&0&0&0&1&-1\\ 
\hdashline[1pt/1.5pt]&&&&&\\[-1em]
1&0&0&-1&0&0\\ 
1&0&0&0&-1&0\\ 
0&1&0&-1&0&0\\ 
0&1&0&0&-1&0\\ 
0&1&0&0&0&-1\\ 
0&0&1&0&-1&0\\ 
0&0&1&0&0&-1\\ 
\hdashline[1pt/1.5pt]\hdashline[1pt/1.5pt]&&&&&\\[-1em]
w_1&w_2&w_3&w_4&w_5&w_6\\ 
\end{array}\right]
\left[\begin{array}{c}
\tau_1 \\ \tau_2 \\ \tau_3 \\ \tau_4 \\ \tau_5 \\ \tau_6
\end{array}\right]
=
\left[\begin{array}{c}
\bar{\Delta}_1 \\ \bar{\Delta}_2 \\ \bar{\Delta}_3 \\ \bar{\Delta}_4 \\ \bar{\Delta}_5 \\ \bar{\Delta}_6 \\
\hdashline[1pt/1.5pt]\hdashline[1pt/1.5pt]\\[-1em]
\bar{\delta}_{12} \\ \bar{\delta}_{23} \\ \bar{\delta}_{45} \\ \bar{\delta}_{56} \\
\hdashline[1pt/1.5pt]\\[-1em]
\bar{\delta}_{14} \\ \bar{\delta}_{15} \\ \bar{\delta}_{24} \\ \bar{\delta}_{25} \\ \bar{\delta}_{26} \\ \bar{\delta}_{35} \\ \bar{\delta}_{36} \\
\hdashline[1pt/1.5pt]\hdashline[1pt/1.5pt]\\[-1em]
\bar{\tau}
\end{array}\right].
\end{equation*}\end{linenomath}
The vertical dashed line separates the matrix coefficients for the strips from the opposing layers ($\Delta E$-layer on the left, $E$-layer on the right). The first horizontal block is formed by a set of equations from Eq.~(\ref{abs_con}). The lowest horizontal block incorporates Eq.~(\ref{add_con}). The middle horizontal block (bounded by double dashed lines) subsumes the constraints from Eq.~(\ref{rel_con}), with the first sub-block corresponding to the coincidences within the same ($\Delta E$ or $E$) layer and the second sub-block to the coincidences between the pairs of strips from opposing layers. The structure of vector $\vec{\x}$ from Eq.~(\ref{vectorized}) is also evident.

In order to perform a weighed fitting, one needs to construct the appropriate weight matrix $\W$. We use a simple diagonal matrix and consider the reliability of specific constraints to be determined by the \textit{sample variance} of the components contributing to $\vec{\x}$, i.e. by their distribution widths (Figs.~\ref{fig2} and \ref{fig3}). The reason is that the \textit{variance of the mean} is sensitive to the amount of statistics (going as \mbox{$N^{-1/2}$} with the number of counts in the unweighted case). On the other hand, the severity of the time offsets between the particular strips is independent of the accumulated statistics, which is better reflected through a sample variance, it being an intrinsic property of the relevant distributions. Relying on Eqs.~(\ref{var_big}) and (\ref{var_small}), while using a notation $\W_{kk}(\cdot)$ in a sense that the $k$-th diagonal element corresponds to the argument in parentheses, we thus define the weighting factors as:
\begin{linenomath}\begin{align}
&\W_{kk}(\bar{\Delta}_s)=1/\var_s\,\Delta,
\label{var_1}\\
&\W_{kk}(\bar{\delta}_{\alpha\beta})=1/\var_{\alpha\beta}\,\delta,
\label{var_2}\\
&\textstyle \W_{kk}(\bar{\tau})=1\,\big/\sum_{s=1}^S w_s^2 \, \var_s\,\Delta.
\label{var_3}
\end{align}\end{linenomath}
For simplicity, in Eq.~(\ref{var_3}) we use the variance estimate based on the coupling of Eqs.~(\ref{abs_con}) and (\ref{add_con}). In the actual implementation one wishes to keep only the statistically significant and otherwise reliable constraints -- based, for example, on the observation of the total number of events ($N_0$ or $N_{\alpha\beta}$) forming the underlying distributions and/or the variance of their means (\mbox{$\var_s\,\bar{\Delta}$} or \mbox{$\var_{\alpha\beta}\,\bar{\delta}$}).

\begin{figure}[t!]
\centering
\includegraphics[width=1\linewidth]{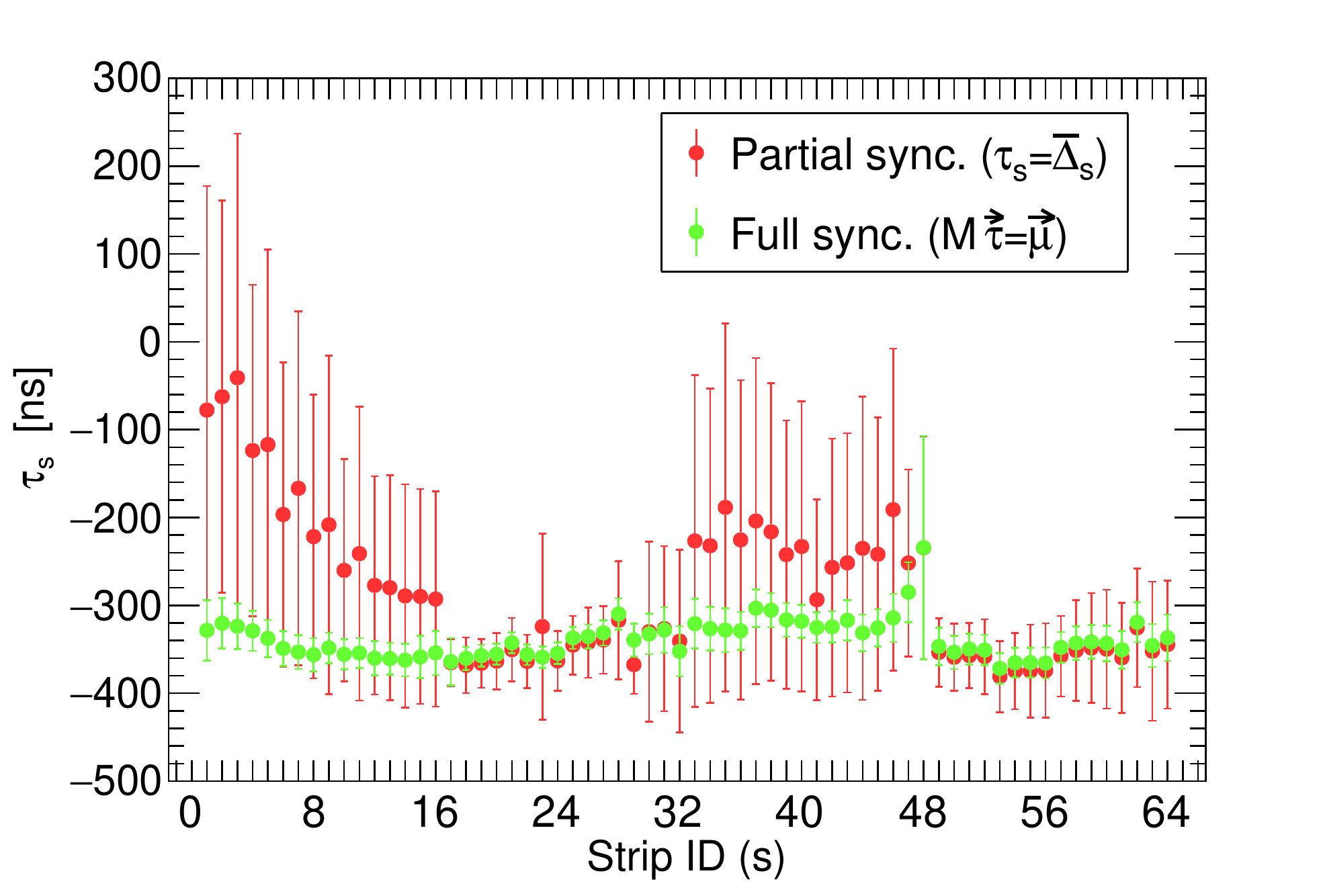}
\pull
\caption{Set of offsets between each particular silicon strip and an external timing device (WCM), obtained either by the partial synchronization from Eq.~(\ref{abs_con}), or the full synchronization accounting for Eqs.~(\ref{abs_con}), (\ref{rel_con}) and (\ref{add_con}).}
\label{fig5}
\end{figure}

Finally, defining the covariance matrix \mbox{$\V=\big(\M^\top \W^{-1}\M\big)^{-1}$}, the solution to the weighted fitting may be expressed as:
\begin{linenomath}\begin{equation}
\vec{\tau}=\V\M^\top \W^{-1}\vec{\x},
\label{solution}
\end{equation}\end{linenomath}
together with accompanying variances \mbox{$\var\,\tau_s=\V_{ss}$},
thus fully resolving a synchronization problem.

\begin{figure}[t!]
\centering
\includegraphics[width=1\linewidth]{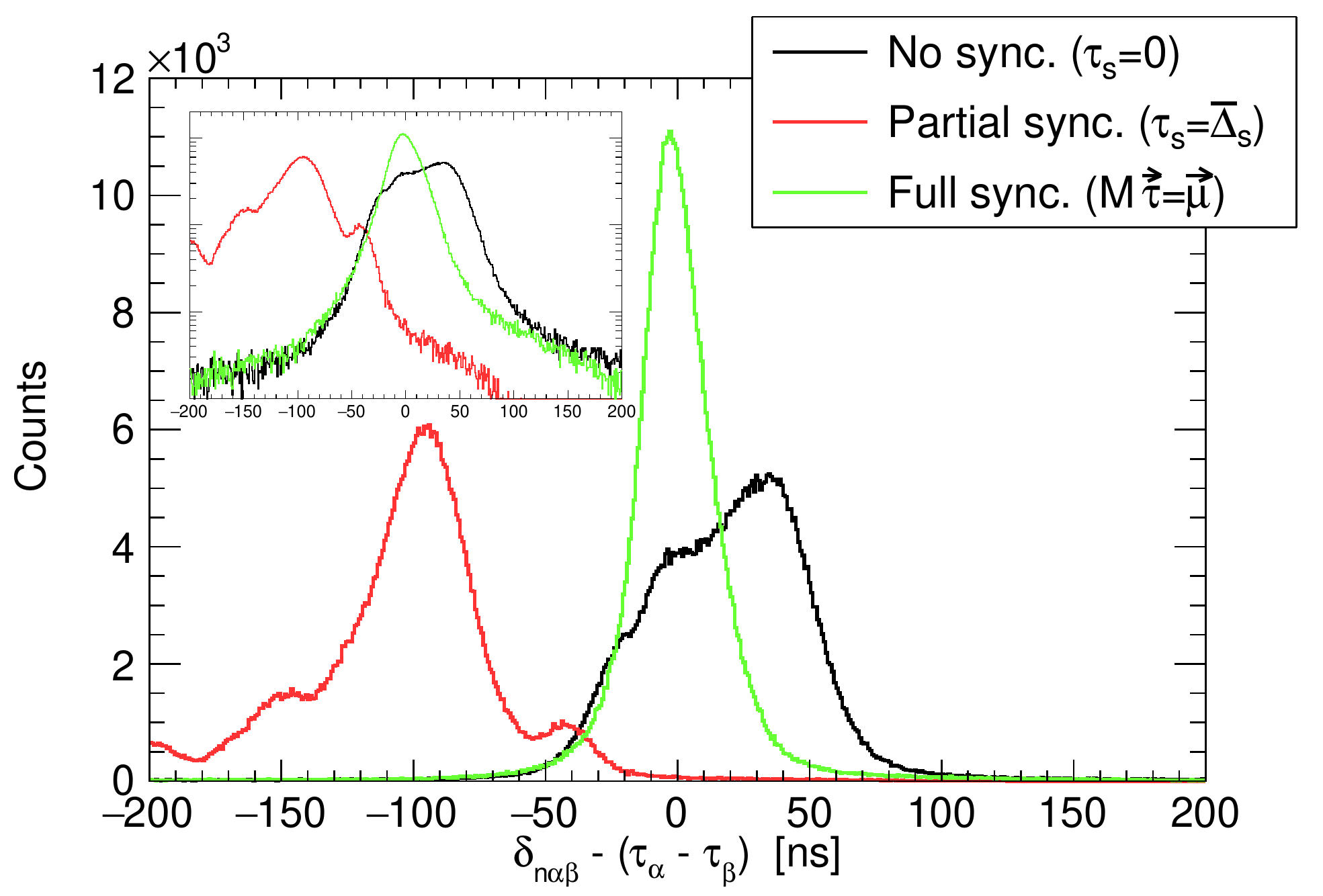}
\pull
\caption{Overall distribution of time differences between coincidental pulses from the measurement of the $^6$Li($n,t$) reaction, registered by any pair of silicon strips. The partial synchronization based solely on Eq.~(\ref{abs_con}) exacerbates the time offsets between the strips. The inset shows the plots in the logarithmic scale, facilitating the selection of $\pm100$~ns as a window width for the identification of coincidental pules.}
\label{fig6}
\end{figure}

Figure~\ref{fig5} shows the difference between the set of offsets obtained by a full synchronization method from Eq.~(\ref{solution}) and those obtained directly from Eq.~(\ref{abs_con}). We refer to the former set of offsets as \textit{fully synchronized} and to the latter as \textit{partially synchronized}. Earlier Figs.~\ref{fig2} and \ref{fig4} also show their respective spectra after being corrected by either of these two sets of offsets. The corrected spectra for each particular strip, contributing to the overall distributions from Fig.~\ref{fig2}, are expected to satisfy the requirement \mbox{$\bar{\Delta}_s-\tau_s\simeq0$} from Eq.~(\ref{abs_con}), either exactly in case of the partial synchronization or as well as possible in case of the full synchronization, considering the complementary requirements from Eqs.~(\ref{rel_con}) and (\ref{add_con}). The corrected spectra from Fig.~\ref{fig4} aim at fulfilling the condition from Eq.~(\ref{lin_ind}). One can clearly observe that -- for purposes of assessing the quality of synchronization, i.e. for discriminating the quality of partial and full synchronization -- the difference between the corrected spectra from Figs.~\ref{fig2} and \ref{fig4} is inconclusive. This is precisely due to the fact that requirements from Eqs.~(\ref{abs_con}) and (\ref{add_con}) are not of sufficient quality \textit{by themselves}, thus having to be complemented by further rigorous constraints from Eq.~(\ref{rel_con}). In that, Fig.~\ref{fig6} shows the overall distribution of time differences between the pairs of coincidental pulses detected by any pair of silicon strips. These spectra clearly show the effects and the quality of two considered types of synchronization, confirming that these are the relevant spectra for the assessment of the synchronization procedure. The uncorrected spectrum is, in essence, an overlap of spectra from Fig.~\ref{fig3}, but taking into account all available pairs of strips. The spectrum obtained by a partial synchronization not only reconfirms that conditions from Eq.~(\ref{abs_con}) are insufficient, but reveals that they are even \textit{inadequate by themselves}, as the offsets between the strips are further exacerbated. The full synchronization expectedly manages to achieve an optimal adjustment between the pulses (i.e. the entire signal waveforms) from the separate strips, as per explicit requirement from Eq.~(\ref{rel_con}). It should be noted that the width of the fully synchronized distribution from Fig.~\ref{fig6} is the basis for selecting the coincidental window width for the identification of coincidental pulses during the analysis of the experimental data from the latest joint measurement of the $^{12}$C($n,p$) and $^{12}$C($n,d$) reactions from n\_TOF. We select it as $\pm100$~ns.

\section{Conclusions}
\label{conclusions}

We presented a simple method for a simultaneous off-line synchronization of the digital data-streams from a multi-channel silicon telescope (SITE). An absolute synchronization is performed relative to the external timing device. A Wall Current Monitor (WCM) -- used for the detection of an instantaneous proton beam from the CERN Proton Synchrotron -- is adopted as the external timing device at n\_TOF. The proton-pulse-coincidental pulses from SITE strips are caused by their response to an intense $\gamma$-flash caused by the instantaneous burst of $\gamma$-rays and ultrarelativistic particles from a spallation process producing the neutron beam. A procedure for obtaining the constraints upon the WCM-relative strip offsets was described, revealing that the minimal necessary subset of these conditions is insufficient for a quality synchronization. Therefore, a relative synchronization between the separate silicon strips was also included into procedure. The relative synchronization is based on observing the coincidental pairs of pulses caused by the detection of charged particles from the neutron induced reactions, alongside the strips' response to the $\gamma$-flash. The full synchronization method was thus expanded to account for the maximum achievable amount of information, in order to constrain the sought offsets as well as possible. Upon completion, a successful synchronization procedure provides an objective evaluation of the coincidental window width to be used in identifying the coincidental pulses during an off-line analysis of the experimental data.


{\color{white}.}

\textbf{Acknowledgements}\\

This work was supported by the Croatian Science Foundation under Project No. 8570.


\appendix

\section{Linear independence of constraints}
\label{linear}

We demonstrate here the linear independence of constraint~(\ref{add_con}) from the set of constraints~(\ref{abs_con}). It should be noted that the requirement of their independence is not related to the invertibility of Eq.~(\ref{vectorized}), as this system of equations is already overdetermined, i.e. carries sufficient information for finding the set of time offsets $\tau_s$. In other words, introducing the linearly dependent term into Eq.~(\ref{vectorized}) does not make the matrix $\M$ singular because $\M$ is larger than the \textit{square} matrix (carrying minimal amount of information) necessary for obtaining $\vec{\tau}$, so that the concept of singularity is not even strictly applicable. Rather, linear dependence must be avoided because it acts as a repeated inclusion of already accounted constraints.

Let us first define a set of terms $\mathcal{T}_s$, obtained by extracting the common denominator from Eq.~(\ref{abs_con}):
\begin{linenomath}\begin{equation}
\textstyle \mathcal{T}_s\equiv \sum_{n=1}^{\ngam} A_{ns} (\Delta_{ns}-\tau_s).
\label{eqA1}
\end{equation}\end{linenomath}
The set of constraints (\ref{abs_con}) is then equal to \mbox{$\mathcal{T}_s\simeq0$}. Let us now carefully write out the left hand side of Eq.~(\ref{lin_ind}), which is equivalent to Eq.~(\ref{add_con}):
\begin{linenomath}\begin{equation}
\sum_{n=1}^{\ngam}(\bar{T}_n-T_n)=\sum_{n=1}^{\ngam}\frac{\sum_{s=1}^S A_{ns} (\Delta_{ns}-\tau_s)}{\sum_{\sigma=1}^S A_{n\sigma}}.
\label{eqA2}
\end{equation}\end{linenomath}
The demonstration of linear (in)dependence boils down to showing if Eq.~(\ref{eqA2}) can be expressed as a linear combination of terms from Eq.~(\ref{eqA1}):
\begin{linenomath}\begin{equation}
\sum_{s=1}^{S}\alpha_s \mathcal{T}_s=\sum_{n=1}^{\ngam}\frac{\sum_{s=1}^S A_{ns} (\Delta_{ns}-\tau_s)}{\sum_{\sigma=1}^S A_{n\sigma}},
\label{eqA3}
\end{equation}\end{linenomath}
that is, if we can find such set of coefficients $\alpha_s$ that Eq.~(\ref{eqA3}) is satisfied. By writing out the left hand side and rearranging the sums from both sides of equation:
\begin{linenomath}\begin{equation}
\sum_{n=1}^{\ngam}\sum_{s=1}^{S}\alpha_s A_{ns} (\Delta_{ns}-\tau_s)=\sum_{n=1}^{\ngam}\sum_{s=1}^{S}\frac{1}{\sum_{\sigma=1}^S A_{n\sigma}}A_{ns} (\Delta_{ns}-\tau_s)
\label{eqA4}
\end{equation}\end{linenomath}
we clearly see that the following should apply:
\begin{linenomath}\begin{equation}
\textstyle \alpha_s=1\,\big/\sum_{\sigma=1}^S A_{n\sigma}.
\label{eqA5}
\end{equation}\end{linenomath}
However, the right hand side of Eq.~(\ref{eqA5}) is a function of the neutron pulse $n$, while the strip-dependent coefficients $\alpha_s$ cannot be! Therefore, a linear dependence from Eq.~(\ref{eqA3}) -- i.e. between Eqs.~(\ref{abs_con}) and (\ref{add_con}) -- cannot  be established.

\end{document}